\documentstyle[aps,pre,twocolumn]{revtex}
\draft
\begin{document}
\def\vec#1{{\bf{#1}}}
\title{Geometric magnetism in classical transport theory}
\author{Jochen Rau}
\address{Max-Planck-Institut f\"ur Physik komplexer Systeme,
Bayreuther Stra{\ss}e 40 Haus 16,
01187 Dresden, Germany}
\date{revised version, May 26, 1997; to appear in Phys. Rev. E}
\maketitle
\begin{abstract}
The effective dynamics of a slow classical system
coupled to a fast chaotic environment is described by means
of a master equation.
We show how this approach permits a very simple
derivation of geometric magnetism.
\end{abstract}
\medskip
\narrowtext
\section{Introduction}
Consider a slow classical system $S$ coupled, through its position,
to a fast classical system $F$.
If the fast motion is chaotic then $F$ effectively acts as
an ``environment'' \cite{environ} which induces friction \cite{friction,jarz2}
and also exerts other, 
non-dissipative reaction forces on the slow system $S$.
As for the non-dissipative reaction,
in the simplest ``adiabatic averaging''
approximation \cite{born_class}
--the classical analogue of the Born-Oppenheimer
approximation \cite{born}-- the fast motion's energy
at given values of the slow coordinates
serves as an external potential for the slow system;
its gradient yields the ``Born-Oppenheimer force.''
In the next approximation beyond this, there is a velocity-dependent
correction which has the form of a magnetic force, and for
which Berry and Robbins coined the name
``geometric magnetism'' \cite{berry_class}.

For the situation considered here, namely a fast chaotic environment,
there are so far two alternative derivations of geometric
magnetism, both
due to Berry and Robbins:
(i) by taking the (non-vanishing) classical limit of the
corresponding quantum result \cite{berry_limit},
where the appearance of geometric magnetism
can be linked to the geometric phase \cite{berry_phase} (whence its name);
or (ii) in a purely classical context, by expanding
the equation of motion for $S$ around the Born-Oppenheimer limit
in powers of the fast/slow time scale ratio, and
identifying the first-order correction \cite{berry_class}.

The multiple-time-scale analysis \cite{ott} used
in the classical derivation of geometric magnetism appears
somewhat reminiscent of the old Chapman-Enskog method to derive
transport equations \cite{balian_book}.
This is not a mere coincidence:
After all,
it should be possible to describe the evolution of
any subsystem (here: $S$) coupled to an environment (here: $F$)
by means of a transport (``master'') equation;
and, 
to be consistent, such a master equation should feature a term
representing geometric magnetism.
It is the purpose of the present paper to show how, indeed,
geometric magnetism arises naturally in a classical
master equation for $S$.
The derivation of geometric magnetism within
this transport theory framework 
will turn out to be surprisingly simple.

The paper is organized as follows.
First I will sketch very briefly how one obtains
transport equations by means of the
Nakajima-Zwanzig projection
technique \cite{nakajima,zwanzig,mori,robertson} 
(Sec. \ref{technique});
for more details the reader is referred to
textbooks \cite{textbooks} and a recent review \cite{rau}.
Then this projection technique is applied
to the situation at hand, namely a slow system $S$ coupled
to a fast chaotic environment $F$, both taken to be classical
(Sec. \ref{app}).
The resulting master equation for $S$ is generally
non-Markovian, yet 
the separation of time scales and chaoticity of the
fast motion permit us to take the Markovian limit.
Along with the Born-Oppenheimer force,
geometric magnetism then appears in a straightforward manner
in the non-dissipative part of the
effective slow dynamics.
Finally, I shall conclude with a brief summary 
and several additional remarks
(Sec. \ref{conclusions}).
\section{Transport equations}\label{technique}
A powerful tool for the derivation of transport equations
is the
Nakajima-Zwanzig projection 
technique \cite{nakajima,zwanzig,mori,robertson,textbooks,rau}.
Its main strength lies in the fact that by
mapping the influence of irrelevant degrees of freedom 
onto --among other features-- a non-local
behavior in time, it opens the way to the 
systematic exploitation of 
separated time scales
and hence serves as a good
starting point for powerful approximations such as
the Markovian and quasistationary limits;
furthermore, it permits one to discern easily the dissipative and
non-dissipative parts of the effective dynamics.

When studying the dynamics of a complex system
away from equilibrium
one typically monitors the evolution of the 
expectation values
\begin{equation}\label{exvalues}
g_a(t):=
\mbox{\bf{(}}\rho(t)|G_a\mbox{\bf{)}}
\end{equation}
of only a very small set of selected
(``relevant'')
observables $\{G_a\}$,
which change according to
\begin{equation}\label{lvn}
\dot g_a(t)=i\mbox{\bf{(}}\rho(t)|{\cal L}G_a\mbox{\bf{)}}
\quad.
\end{equation}
Here the meanings of $\rho(t)$, $G_a$ and $\mbox{\bf{(}}|\mbox{\bf{)}}$, as
well as of the scalar product
$\langle;\rangle_{\rho}$ which we shall use later,
depend on whether the system under consideration
is quantum or classical;
they are summarized in Table I.   %\ref{meaning}.
The ``Liouvillian'' ${\cal L}$ takes the commutator with the
Hamilton operator $\hat{H}$,
\begin{equation}
i{\cal L}=(i/\hbar)\,[\hat{H},*]
\quad,
\end{equation}
for a quantum system, or generates a Lie dragging in the
direction of the Hamiltonian vector $X_H$,
\begin{equation}
i{\cal L}=\pounds_{X_H}
\quad,
\label{L_class}
\end{equation}
for a classical system \cite{arnold}, respectively.
For simplicity we assume that the Hamiltonian and hence the
Liouvillian, as well as 
the relevant observables,
are not explicitly time-dependent.

The right-hand side of the equation of motion (\ref{lvn}) 
will generally depend not just on the selected, but also on
all other (``irrelevant'') degrees of freedom.
In order to eliminate these and hence obtain a closed
``transport equation'' for the $\{g_a(t)\}$,
one employs a suitable projection operator 
which projects
arbitrary observables onto the
subspace spanned by $1$ and the
relevant observables $\{G_a\}$.
The projector may depend 
on the current expectation values of the relevant observables
and thus vary in
time, ${\cal P}(t)\equiv {\cal P}[g_a(t)]$,
and is assumed to have the three properties
(i) ${\cal P}(t)^2={\cal P}(t)$; 
(ii) ${\cal P}(t) A= A$ if and only if
$A\in {\rm span}\{1,{G_a}\}$;
and (iii)
\begin{equation}
\left(
{\rho(t)}\left|{{d}{\cal P}(t)\over{d} t}A\right.
\right)
=0
\quad\forall\,\,\rho(t),A
\quad.
\label{technical}
\end{equation}
Its complement
is denoted by ${\cal Q}(t):=1-{\cal P}(t)$.
One further defines an operator
${\cal T}(t',t)$ (again in the space of
observables) by
\begin{equation}
\label{irrel:evolution}
{\partial\over\partial t'}{\cal T}(t',t)=
-{i}\,{\cal Q}(t'){\cal L}{\cal Q}(t')
{\cal T}(t',t) 
\end{equation}
with initial condition
${\cal T}(t,t)=1$,
which may be pictured as describing
the evolution of the system's {\em irrelevant} degrees of freedom.
The equation of motion
for the selected expectation values $\{g_a(t)\}$
can then be cast into the --still exact-- form
\begin{eqnarray}
\dot g_a (t)
&=& 
{i}\mbox{\bf{(}}\rho(t)|{\cal P}(t){\cal L}{G_a}\mbox{\bf{)}}
\nonumber \\
&-&
\int_{0}^t{d} t'\,
\mbox{\bf{(}}\rho(t')|{\cal P}(t'){\cal L}{\cal Q}(t')
{\cal T}(t',t){\cal Q}(t){\cal L}{G_a}\mbox{\bf{)}}
\nonumber \\
&+& 
{i}\mbox{\bf{(}}\rho(0)|{\cal Q}(0)
{\cal T}(0,t){\cal Q}(t){\cal L}{G_a}\mbox{\bf{)}}
\quad,
\label{robertson_2}
\end{eqnarray}
for any time $t\ge 0$.
This constitutes the desired closed 
system of (possibly nonlinear) coupled integro-differential equations
for the selected expectation values 
$\{g_a(t)\}$ {\em if and only if}
$\mbox{\bf{(}}\rho(0)|{\cal Q}(0)$ and with it the last
(``residual force'') term vanishes.

In many practical applications the initial state $\rho(0)$ is
not known exactly but
characterized solely by the initial expectation values
$\{g_a(0)\}$
of the relevant observables.
From this insufficient information one 
generally constructs that distribution
which maximizes the entropy
$S[\rho]:=-k\mbox{\bf{(}}\rho|\ln\rho\mbox{\bf{)}}$ and hence can be
considered
``least biased'' or ``maximally non-committal'' 
with regard to the unmonitored degrees of freedom:
It is the
generalized canonical state
\begin{equation}
\rho(0)=Z(0)^{-1}\exp[- \lambda^a(0) G_a]
\quad,
\label{initial_state}
\end{equation}
with summation over $a$ implied (Einstein convention),
partition function
\begin{equation}
Z(0):=\mbox{\bf{(}}1|\exp[- \lambda^a(0) G_a]\mbox{\bf{)}}
\end{equation}
and the Lagrange parameters $\{\lambda^a(0)\}$ adjusted such
as to yield the correct $\{g_a(0)\}$.

In an analogous fashion one defines a ``relevant part'' 
\begin{equation}\label{gen_canon}
\rho_{\rm rel}(t):=Z(t)^{-1}\exp[
- \lambda^a(t) G_a]
\end{equation}
of the exact state $\rho(t)$ at {\em all} times $t$, 
where $\rho_{\rm rel}(0)=\rho(0)$ but generally
$\rho_{\rm rel}(t)\ne \rho(t)$ for $t>0$.
There exists a unique time-dependent
projector ${\cal P}_{\rm R}(t)$,
namely the projector 
orthogonal with respect to the time-dependent scalar product
$\langle ;\rangle_{\rho_{\rm rel}(t)}$,
which has all
required properties (i)--(iii) and, moreover,
yields
\begin{equation}
\mbox{\bf{(}}\rho(t)|{\cal P}_{\rm R}(t)=\mbox{\bf{(}}\rho_{\rm rel}(t)|
\end{equation}
at all times.
This special
choice, originally proposed by 
Robertson \cite{robertson,newfoot},
has the important advantage that for initial states of the
form (\ref{initial_state})
it ensures $\mbox{\bf{(}}\rho(0)|{\cal Q}_{\rm R}(0)=0$ and
so renders the transport equation (\ref{robertson_2}) closed.
We shall use the Robertson projector throughout the remainder of the paper
(and, for brevity, immediately drop the subscript `R').

One principal feature of the transport equation (\ref{robertson_2}) 
is that it is non-Markovian:
Future expectation values of the selected observables are
predicted on the basis of both their present values
and their past history.
There are two distinct time scales:
(i) the scale $\tau_{\rm rel}$ --or several scales 
$\{\tau_{\rm rel}^{i}\}$-- on which the 
selected expectation values $\{g_a(t)\}$
evolve;
and (ii) the ``memory time'' $\tau_{\rm mem}$ which
characterizes the length of the time interval that contributes
significantly to the memory integral. 
Only if this memory time is small compared to
the typical time scale on which the selected observables evolve,
$\tau_{\rm mem}\ll\tau_{\rm rel}$, 
can memory effects be neglected and predictions for the
selected observables be based solely on their present values.
One may then assume that
in the memory term $g_a(t')\approx g_a(t)$ and hence
replace
\begin{eqnarray}
{\cal P}[g_a(t')]
&\to&
{\cal P}[g_a(t)]
\quad,
\nonumber \\
\mbox{\bf{(}}\rho(t')|{\cal P}(t')
&\to&
\mbox{\bf{(}}\rho(t)|{\cal P}(t)
\quad,
\nonumber \\
{\cal T}(t',t)
&\to&
\exp[i{\cal Q}(t){\cal L}{\cal Q}(t)\cdot(t-t')]
\end{eqnarray}
(Markovian limit).
Furthermore,
at times $t\gg\tau_{\rm mem}$ it no longer matters 
for the dynamics when exactly
the evolution started, and hence in Eq. (\ref{robertson_2})
the integration over the
system's history may just as well extend from $-\infty$ to $t$,
rather than from $0$ to $t$ (quasistationary limit) \cite{quasi}.
In the Markovian and quasistationary limits
the equation of motion
simplifies to
\begin{eqnarray}
\dot{g_a}(t)
&=&
{i}\mbox{\bf{(}}\rho(t)|{\cal L}_{\rm rel}(t)G_a\mbox{\bf{)}}
\nonumber \\
&&-\pi
\mbox{\bf{(}}\rho(t)|{\cal P}(t){\cal L}{\cal Q}(t)
\delta({\cal Q}(t){\cal L}{\cal Q}(t))
{\cal Q}(t){\cal L}G_a\mbox{\bf{)}}
\label{reformulated}
\end{eqnarray}
modulo residual force, where 
\begin{eqnarray}
{\cal L}_{\rm rel}(t)
&=&
{\cal P}(t){\cal L}{\cal P}(t)
\nonumber \\
&&+
{i\over2}
\int_0^\infty\! d\tau\,
{\cal P}(t){\cal L}[{\cal L}(t;\tau)-{\cal L}(t;-\tau)]{\cal P}(t)
\label{L_rel}
\end{eqnarray}
with
\begin{equation}
{\cal L}(t;\tau):=
\exp[i{\cal Q}(t){\cal L}{\cal Q}(t)\tau]
\,{\cal L}
\label{L_rotated}
\end{equation}
denotes a --possibly time-dependent-- effective Liouvillian
for the relevant observables.

Provided the evolution operator ${\cal T}$ is unitary with
respect to the scalar product
$\langle;\rangle_{\rho_{\rm rel}(t)}$
then the first term in the Markovian transport
equation (\ref{reformulated}) is non-dissipative.
In this case dissipation stems entirely from
the second term, which yields a non-negative
entropy growth rate
\begin{eqnarray}
\dot S[\rho_{\rm rel}(t)]&=&k\,\lambda^a(t)\dot g_a(t)
\nonumber \\
&=&
k\,\pi \langle{\cal Q}{\cal L}\lambda^b G_b;
\delta({\cal Q}{\cal L}{\cal Q}){\cal Q}{\cal L}\lambda^a G_a
\rangle_{\rho_{\rm rel}} 
\ge 0
\end{eqnarray}
($H$-theorem).
\section{Effective force caused by a fast chaotic environment}\label{app}
We now apply the above general results to a slow system $S$ coupled
to a fast chaotic, but not necessarily macroscopic, environment $F$.
Both $S$ and $F$ are treated classically, and their state
described in a phase space with canonical coordinates
${Z}=\{\vec{Q},\vec{P}\}$ pertaining to 
$S$ and ${z}=\{\vec{q},\vec{p}\}$ pertaining
to $F$, respectively.
The full Hamilton function for the combined system $S\times F$ 
is taken to be of the form
\begin{equation}
{H}({Z},{z})={H}_{\rm S}({Z}) + {h}(\vec{Q},z)
\quad,
\end{equation}
where $H_{\rm S}$
governs
the free dynamics of the system $S$, 
and $h$
describes both the
coupling --through the slow position $\vec{Q}$--
of $S$ to the environment 
and the internal dynamics of the latter.
Associated with the Hamilton function is a Liouvillian (\ref{L_class})
which we decompose 
\begin{equation}
\pounds_{X_H}=\pounds_{X_{H,Z}}+\pounds_{X_{H,z}}
\end{equation}
into a part dragging along the slow coordinates,
\begin{equation}
{X_{H,Z}}=
\sum_i\left(V^i {\partial\over\partial Q^i} -
({\partial_i H_S}+{\partial_i h})
{\partial\over\partial P_i}
\right)
\label{L_slow}
\end{equation}
with slow velocity $V^i=\partial H_S/\partial P_i$ and
$\partial_i:={\partial/\partial Q^i}$,
and a part dragging along the fast coordinates,
\begin{equation}
{X_{H,z}}=
\sum_k\left({\partial h\over\partial p_k}\, {\partial\over\partial q^k} -
{\partial h\over\partial q^k}\,{\partial\over\partial p_k}
\right)
\quad.
\end{equation}

At $t=0$ and hence, due to energy conservation, at all times
the combined system $S\times F$ is assumed to have
a sharp total energy $E$.
For the purposes of the Nakajima-Zwanzig projection technique
all observables pertaining to $S$, as well as the total
energy which is a constant of the motion, are taken to be relevant;
while the internal degrees of freedom
of the environment and system-environment correlations are deemed irrelevant.
This gives rise to a
time-independent representation of the Robertson projector,
\begin{eqnarray}
{\cal P}A&=&
{1\over\partial_E\Omega}
\int dz\,\delta({H}-E)\,{A}
\nonumber \\
&=:&
\langle A\rangle_E
\end{eqnarray}
for any observable $A$.
Here $\partial_E:=\partial/\partial E$ and
\begin{equation}
\Omega:=
\int dz\,\theta({H}-E)
\quad;
\end{equation}
its derivative $\partial_E\Omega$ may be interpreted as the 
surface of the microcanonical energy shell.

The slow system's effective dynamics must be described 
with a transport equation of the form (\ref{robertson_2}),
which in general is non-Markovian and includes a residual force.
Only if we take the initial state of the environment
to be microcanonical, i.e.,
\begin{equation}
\rho(0)=\rho_{\rm S}(0)\times{1\over\partial_E\Omega}
\delta(H-E)
\label{microcan}
\end{equation}
where $\rho_{\rm S}(0)$
denotes the (arbitrary) initial state of $S$,
then the residual force term vanishes \cite{fluc_diss}.
This assumption of a microcanonical distribution and the
resultant omission of the residual force term 
amount to averaging over an entire ensemble of fast chaotic systems.
However, even if the slow system is coupled to
a single fast chaotic system the 
transport equation without residual force will describe the
main global feature of the slow dynamics; 
the residual force
causes only fluctuations around the average trajectory
\cite{berry_example}.

The separation of time scales and chaoticity of the
fast motion permit us to take the Markovian and
quasistationary limits.
Moreover, we focus on the non-dissipative part of the
slow dynamics.
The latter is governed by the effective Liouvillian (\ref{L_rel}),
which immediately yields the (Heisenberg-picture)
equation of motion
\begin{eqnarray}
\dot{G}
&=&
\langle \pounds_{X_{H,Z}}{G} \rangle_E
\nonumber \\
&&+ 
{1\over2}\int_0^\infty d\tau\,
\langle \pounds_{X_{H,Z}}
[ \pounds_{X_{H,Z}(\tau)}- 
\pounds_{X_{H,Z}(-\tau)}]
{G}\rangle_E
\label{slow_eom}
\end{eqnarray}
for an arbitrary slow observable $G(Z)$ \cite{not_hamiltonian}.
Here we have used $\pounds_{X_{H,z}}{\cal P}
={\cal P}\pounds_{X_{H,z}}=0$ to replace
$\pounds_{X_{H}}$ by $\pounds_{X_{H,Z}}$, and defined
the ``rotated'' Hamiltonian vector
${X_{H,Z}(\tau)}$ as in (\ref{L_slow}) but
with components dragged along the fast coordinates:
\begin{equation}
{\partial_i h}\to
({\partial_i h})_\tau :=
\exp[\tau\pounds_{X_{H,z}}]\,
({\partial_i h})
\quad.
\end{equation}
Upon choosing $G={\vec{P}}$ we obtain
an effective force,
with components
\begin{eqnarray}
\dot{P}_i
&=&
- \langle \partial_i H \rangle_E
- {1\over2} \sum_j V^j \!\int_0^\infty \!d\tau\,
\langle \partial_j
[(\partial_i h)_\tau - (\partial_i h)_{-\tau}]
\rangle_E
\nonumber \\
&=&
F_i^{\rm BO} + F_i^{\rm geo}
\quad .
\end{eqnarray}
The first term constitutes the usual Born-Oppenheimer 
force;
while the second (integral) term gives rise to geometric magnetism:
For an arbitrary function $A(Z,z)$ it is
\begin{eqnarray}
\langle\partial_j A\rangle_E
&=&
\partial_j\langle A\rangle_E + \langle\partial_j H\rangle_E
\partial_E \langle A\rangle_E
\nonumber \\
&&+
{1\over\partial_E \Omega}\partial_E \left[\partial_E \Omega
\cdot\langle A\,\tilde{\partial}_j H\rangle_E\right]
\, ,
\label{identity}
\end{eqnarray}
where $\tilde{\partial}_j H:={\partial}_j H-
\langle{\partial}_j H \rangle_E$.
This identity, together with 
$\tilde{\partial}_j H=\tilde{\partial}_j h$,
\begin{equation}
\langle(\partial_i h)_\tau \rangle_E =
\langle(\partial_i h)_{-\tau} \rangle_E
\end{equation}
and
\begin{equation}
\langle (\tilde{\partial}_i h)_{-\tau}\tilde{\partial}_j h\rangle_E
= 
\langle (\tilde{\partial}_j h)_{\tau}\tilde{\partial}_i h\rangle_E
\quad,
\end{equation}
yields 
\begin{equation}
F_i^{\rm geo} = \sum_j B_{ij} V^j
\end{equation}
with an antisymmetric matrix (``magnetic field'')
\begin{equation}
B_{ij}=
-{1\over2\partial_E \Omega}\partial_E \!\left[\partial_E \Omega
\!\int_0^\infty \!d\tau
\langle (\tilde{\partial}_i h)_{\tau}\tilde{\partial}_j h -
(\tilde{\partial}_j h)_{\tau}\tilde{\partial}_i h\rangle_E
\right]
\label{field}
\end{equation}
in agreement with the result by Berry and
Robbins \cite{berry_class,berry_limit}.
\section{Conclusion}\label{conclusions}
We have succeeded in deriving geometric magnetism within
the framework of classical transport theory:
Starting from the general formula (\ref{L_rel}),
the derivation turned out to be surprisingly simple.
This means that treating the fast chaotic system as an
``environment'' for the slow system and describing the
dynamics of the latter with a master equation
is a consistent and useful physical picture.
More generally, it shows that methods from transport theory are
not limited to the description of macroscopic systems but 
apply as well to low-dimensional chaos.

As a by-product we have obtained an equation of motion for
arbitrary slow observables (Eq. (\ref{slow_eom}))
which is coordinate-free and applies to any form
of the microscopic Hamilton function:
It could serve as the starting point for interesting generalizations
of the model Hamiltonian considered here.
Higher-order corrections to the Born-Oppenheimer force and
geometric magnetism will presumably have to take into
account memory effects; an example for this is
Jarzynski's force \cite{jarzynski}.
Here, too, transport theory with its non-Markovian evolution
equation (\ref{robertson_2}) will furnish a good starting point
for the systematic study of non-Markovian corrections.

Finally, transport theory adds a somewhat new perspective
to Berry and Robbins' observation that geometric magnetism is the
``antisymmetric cousin of friction'' \cite{berry_class}.
In our formulation the memory term in Eq. (\ref{robertson_2}) 
gives rise, upon taking the Markovian and quasistationary limits,
to both 
dissipative and non-dissipative parts of the dynamics; they
appear as the parts symmetrized and antisymmetrized,
resp., with respect to the history integration variable $\tau=(t-t')$.
When evaluated for our model Hamiltonian 
and the special choice $G=\vec{P}$
these two parts translate 
into effective forces proportional to the slow velocity,
one with a symmetric matrix of coefficients (friction:
not considered in this paper),
the other with an antisymmetric matrix
(geometric magnetism: Eq. (\ref{field})).
\acknowledgements
I thank T. Dittrich for helpful discussions.
%

%
%
% table %%%%%
%
{
\begin{table}
\label{meaning}
\caption{Various symbols used in transport theory}
\begin{tabular}{c|c|c}
generic symbol&
quantum&
classical
\\
\hline
state $\rho$ &
statistical op. $\hat{\rho}$ &
phase space dist. $\rho(\vec{\pi})$
\\ 
observable $G_a$ &
Hermitian op. $\hat{G}_a$ &
real function $G_a(\vec{\pi})$
\\ 
$\mbox{\bf{(}}A|B\mbox{\bf{)}}$ &
$\mbox{tr}(\hat{A}^\dagger \hat{B})$ &
$\int d\vec{\pi}\, A^*(\vec{\pi}) B(\vec{\pi})$
\\ 
$\langle A;B\rangle_\rho$ &
$\int_0^1 d\mu\,\mbox{tr}\left[\hat{\rho}^\mu
\hat{A}^\dagger \hat{\rho}^{1-\mu}\hat{B}\right]$ &
$\int d\vec{\pi}\, \rho(\vec{\pi}) A^*(\vec{\pi}) B(\vec{\pi})$
\\ 
\end{tabular}
\end{table}
}

\begin{references}
\bibitem{environ}
not to be confused with a macroscopic heat bath
\bibitem{quasi}
The quasistationary limit is not justified if the system 
exhibits a finite recurrence time $\tau_{\rm rec}$,
as is the case for a finite quantum system with discrete spectrum.
The history integration
in Eq. (\ref{robertson_2}) must then remain limited to a time interval
much shorter than $\tau_{\rm rec}$, and
the resultant transport equation will only be valid at intermediate times
$\tau_{\rm mem}\ll t\ll\tau_{\rm rec}$.
Taking the quasistationary limit would lead to 
paradoxical results such as the disappearance of friction
upon quantization of a finite system.
For a discussion of this apparent ``quantum/classical discordance'' 
and a consideration of the time scales involved, 
see Ref. \cite{berry_class}.
\bibitem{fluc_diss}
The residual force should not be confused with the stochastic
force that enters into the
fluctuation-dissipation theorem \cite{jarz2}, so
a vanishing residual force does not imply that there is no friction.
Friction stems from the second term in Eq. (\ref{reformulated}).
\bibitem{not_hamiltonian}
The effective dynamics need not necessarily be hamiltonian:
There need not exist $H_{\rm rel}$ such that
$i{\cal L}_{\rm rel}=\pounds_{X_{H_{\rm rel}}}$.
\end{references}
\end{document}